\documentclass[11pt]{article}

\usepackage[final]{acl}

\usepackage{times}
\usepackage{latexsym}

\usepackage[T1]{fontenc}

\usepackage[utf8]{inputenc}

\usepackage{microtype}

\usepackage{inconsolata}

\usepackage{graphicx}

\usepackage{amsmath}
\usepackage{amsfonts}
\usepackage{subfigure}
\usepackage{booktabs}
\usepackage{xcolor}
\usepackage{multirow}

\title{STK-Adapter: Incorporating Evolving Graph and Event Chain for Temporal Knowledge Graph Extrapolation}

\author{
 \textbf{Shuyuan Zhao\textsuperscript{1,3}},
 \textbf{Wei Chen\textsuperscript{2}}\thanks{Corresponding author.},
 \textbf{Weijie Zhang\textsuperscript{1,3}},
 \textbf{Xinrui Hou\textsuperscript{1,3}},
 \textbf{Junfeng Shen\textsuperscript{1,3}},
 \\
 \textbf{Boyan Shi\textsuperscript{1,3}},
 \textbf{Shengnan Guo\textsuperscript{1,3}},
 \textbf{Youfang Lin\textsuperscript{1,3}},
  \textbf{Huaiyu Wan\textsuperscript{1,3}}
  \\
 \textsuperscript{1}School of Computer Science and Technology, Beijing Jiaotong University, China\\
  \textsuperscript{2}Guangxi Key Lab of Trusted Software, Guilin University of Electronic Technology, China\\
 \textsuperscript{3}Beijing Key Laboratory of Traffic Data Mining and Embodied Intelligence, China
\\
 \small{
    \{sy\_zhao, hywan\}@bjtu.edu.cn, \textbf{Correspondence:} w\_chen@guet.edu.cn
 }
}

\begin{document}
\maketitle
\begin{abstract}
Temporal Knowledge Graph (TKG)  extrapolation aims to predict future events based on historical facts.
Recent studies have attempted to enhance TKG extrapolation by integrating TKG's evolving structural representations and textual event chains into Large Language Models (LLMs).
Yet, two main challenges limit these approaches:
(1) The loss of essential spatial-temporal information due to shallow alignment between TKG's graph evolving structural representation and the LLM's semantic space, and (2) the progressive dilution of the TKG's evolving structural features during LLM fine-tuning.
To address these challenges, we propose the \textbf{S}patial-\textbf{T}emporal \textbf{K}nowledge\textbf{ Adapter} (\textbf{STK-Adapter}), 
which flexibly integrates the evolving graph encoder and the LLM to facilitate TKG reasoning.
In STK-Adapter, a Spatial-Temporal MoE is designed to capture spatial structures and temporal patterns inherent in TKGs.
An Event-Aware MoE is employed to model intricate temporal semantic dependencies within event chains. 
In addition, a Cross-Modality Alignment MoE is proposed to facilitate deep cross-modality alignment by TKG-guided attention experts.
Extensive experiments on benchmark datasets demonstrate that STK-Adapter significantly outperforms state-of-the-art methods and exhibits strong generalization capabilities in cross-dataset tasks.
The code is available at \url{https://github.com/Zhaoshuyuan0246/STK-Adapter}.
\end{abstract}

\section{Introduction}
Temporal Knowledge Graphs (TKGs) store dynamic facts in the real world.
Each fact, or event, in TKGs is formally represented as a structured quadruple: (subject, relation, object, timestamp).
TKG Extrapolation aims to predict future events based on historical facts, playing a vital role in critical domains such as financial risk management, recommendation systems, and medical assistance \cite{zhaoshuyuan, yaochang, chenwei_2, yinzihao}.

\begin{figure}[t]
  \centering
    \includegraphics[width=1\linewidth]{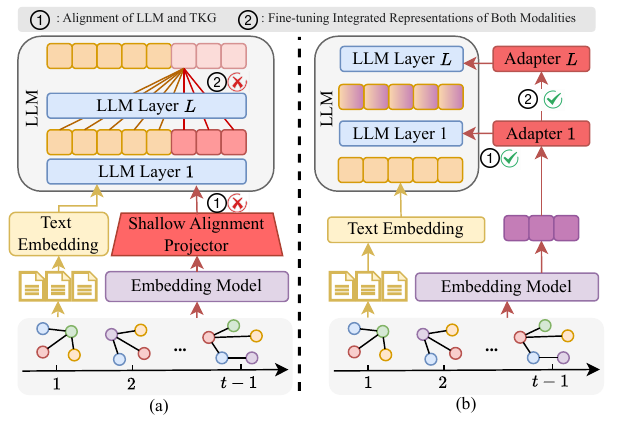}
  \caption{Limitations of existing approaches for integrating TKG embeddings with LLMs, and the intuition behind our proposed STK-Adapter.
  (a) The shallow projector hinders the alignment of TKG embeddings (red blocks) with text embeddings (yellow blocks), and the evolving structural information inherent in TKG is progressively diluted across LLM layers during fine-tuning.
  (b) Our STK-Adapter integrates TKG embeddings (purple blocks) at each layer, facilitating deep and progressive alignment with the LLM's semantic space.}
\label{fig1}
\end{figure}

Previous research on TKG extrapolation, such as TiRGN \cite{TiRGN} and LogCL \cite{LogCL}, focuses on capturing spatial-temporal dependencies among events by modeling TKG snapshot sequences.
Despite their effectiveness, these methods embed events into latent representation spaces, which lose the textual semantic aspects of events in TKGs \cite{GenTKG}.
%
With the emergence of Large Language Models (LLMs), several approaches have leveraged their powerful semantic reasoning capabilities for TKG extrapolation \cite{GenTKG, lama-2-7b-CoH}.
The LLM-based methods linearize structured TKGs into textual event chains via various retrieval strategies and employ them for instruction fine-tuning. 
Yet, the linearization process hinders LLMs from fully preserving and modeling the evolving structural features inherent in TKGs \cite{LLMuseGNN}.
Recent studies, such as GenTKGQA \cite{GenTKGQA} and TGL-LLM \cite{TGL-LLM}, attempt to integrate TKG's evolving structural representations with textual event chains into LLMs.
By fine-tuning LLMs to align graph embeddings with their semantic space, these approaches jointly leverage spatial-temporal structural signals and the inherent semantic reasoning capabilities of LLMs, improving extrapolation performance.
Nevertheless, these methods suffer from two critical limitations:
\textbf{(1) Shallow alignment of TKG's evolving structural representations with the LLM's semantic space results in the loss of essential spatial-temporal information.}
During the alignment, previous methods employ simplistic one-shot projection techniques (e.g., Multi-Layer Perceptron) to map TKG's evolving structural representations into tokens within the LLM's semantic space, as illustrated in Figure \ref{fig1}(a). 
This superficial alignment fails to fully preserve TKG's spatial-temporal information.
\textbf{(2) The TKG's evolving structural features are progressively diluted within the LLM during the fine-tuning.}
LLMs are inherently optimized for next-token prediction, which biases their hidden states toward textual semantics.
As shown in Figure~\ref{fig1}(a), during fine-tuning, LLMs prioritize textual semantic representations, while the evolving structural features of TKGs are progressively diluted across layers, limiting the LLM's capacity to model temporal graph structures.

To address these limitations, we propose \textbf{S}patial-\textbf{T}emporal \textbf{K}nowledge \textbf{Adapter} (\textbf{STK-Adapter}), which establishes dedicated processing pathways in each LLM layer to progressively capture TKG's evolving structural representations and textual event chains.
Inspired by the Mixture-of-Experts (MoE), widely adopted in LLMs to model complex and heterogeneous data \cite{fusemoe}, the STK-Adapter involves three MoE modules with independent expert pools: a Spatial-Temporal MoE (ST-MoE), an Event-Aware MoE (EA-MoE), and a Cross-Modality Alignment MoE (CMA-MoE).
The ST-MoE employs specialized experts to capture diverse spatial structures and temporal patterns inherent in TKGs.
The EA-MoE guides the LLM to learn complex temporal semantic dependencies in event chains through an event-aware router.
The CMA-MoE employs TKG-guided attention experts to facilitate deep, cross-modality alignment.
STK-Adapter ensures that both the evolving structure of TKGs and the semantic context of event chains are effectively leveraged by the LLM for TKG extrapolation.

The contributions of our work are as follows:
\begin{itemize}
    \item  We propose a novel STK-Adapter that flexibly integrates an evolving graph encoder with LLMs to facilitate TKG extrapolation, achieving a synergistic advantage between evolving graph structures and textual semantics.

    \item 
    We design three MoE modules to establish dedicated pathways for capturing spatial-temporal features in TKGs,  modeling temporal semantic dependencies within event chains, and achieving deep cross-modality alignment, ensuring TKG's evolving structures and event chains are effectively captured and integrated in the LLM for TKG extrapolation.
    

    \item Extensive experiments demonstrate that STK-Adapter significantly outperforms state-of-the-art methods on benchmark datasets and exhibits strong generalization capabilities in cross-dataset tasks.
\end{itemize}

\begin{figure*}[htbp]
    \centering
    \includegraphics[width=1 \linewidth]{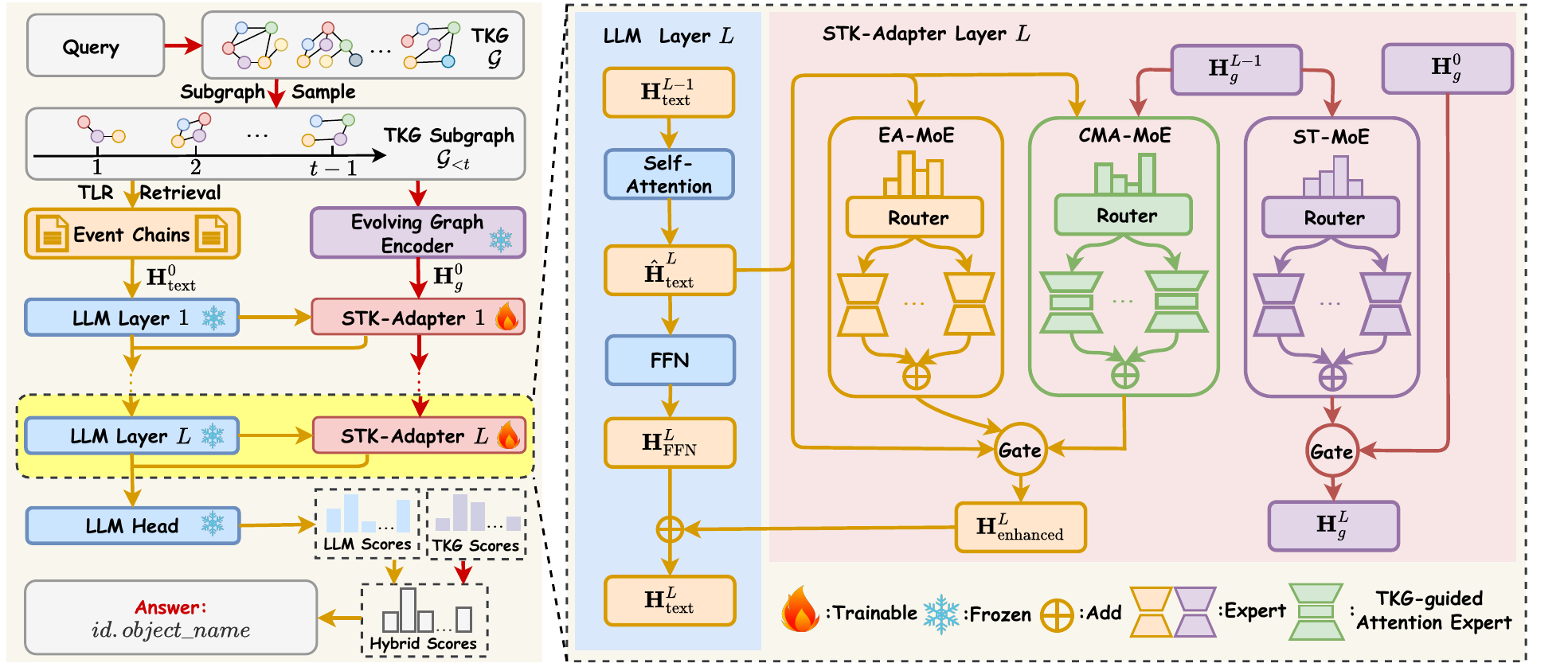}
    \caption{The overall architecture of STK-Adapter consists of three MoEs: the ST-MoE, the EA-MoE, and the CMA-MoE.}
    \label{fig2}
\end{figure*}

\section{Related Work}
\subsection{Temporal Knowledge Graph Extrapolation}
Temporal Knowledge Graph Extrapolation aims to predict future events by leveraging historical information in TKGs.
Embedding-based methods employ deep neural networks to model historical events in the latent space \cite{zhiyu, chenwei_1}. 
For example, REGCN \cite{RE-GCN} captures both spatial and temporal dependencies across TKG snapshots.
With the emergence of LLMs, some methods have explored their application to TKG extrapolation \cite{wan_3, wan_4}.
Methods such as CoH \cite{lama-2-7b-CoH} linearize TKGs into event chains and fine-tune LLMs for prediction.
Nevertheless, linearization inevitably discards the intricate topological structures of TKGs \cite{LLMuseGNN}.
To mitigate this issue, recent studies inject TKG representations into LLMs via projector modules. 
For instance, GenTKGQA \cite{GenTKGQA} and TGL-LLM \cite{TGL-LLM} align TKG's evolving structural representations with the LLM’s semantic space via a linear layer or MLPs.
However, these approaches often result in shallow alignment and progressive dilution of the TKG's evolving structures.
In contrast, our work deeply integrates TKG representations and event chains with LLMs, enabling LLMs to jointly capture the evolving graph structures and textual semantics, facilitating efficient TKG extrapolation.

\subsection{Mixture of Experts}
A Mixture-of-Experts consists of a router and multiple expert networks, where the router adaptively activates a subset of experts for each input.
By assigning inputs to specialized experts, MoE enables sparse computation while significantly enhancing model capacity and performance \cite{MoE, MoE_Structure}.
Due to its ability to handle heterogeneous inputs through expert specialization, MoE has been widely adopted in multi-modal applications \cite{fusemoe}.


Recently, MoE has been applied to Parameter-Efficient Fine-Tuning (PEFT) of LLMs by integrating multiple low-rank experts \cite{Adapter, lora} in parallel into the Feed-Forward Network (FFN) layers  \cite{Aadamix, More, wuhao, HMoRA}. 
Such a design facilitates specialized feature extraction while preserving LLM's foundational knowledge, alleviating catastrophic forgetting \cite{Loramoe}. 
Inspired by these advances, we propose STK-Adapter, which incorporates three low-rank MoE modules for capturing spatial-temporal features in TKGs, modeling temporal semantic dependencies within event chains, and achieving deep
cross-modality alignment.

\section{Preliminaries}
A Temporal Knowledge Graph, denoted as $\mathcal{G}$, is formally defined as a sequence of TKG snapshots ordered by timestamps: $\mathcal{G} = \{ \mathcal{G}_1, \mathcal{G}_2, \dots, \mathcal{G}_{|\mathcal{T}|} \}$. 
Each snapshot $\mathcal{G}_t = (\mathcal{E},\mathcal{R},\mathcal{F}_t)$ is a directed multi-relational graph with temporal attributes, recording the set of events $\mathcal{F}_t$ at timestamp $t \in \mathcal{T}$. 
Each event is represented as a quadruple $(s, r, o, t)$, where the subject entity $s \in \mathcal{E}$ and the object entity $o \in \mathcal{E}$ are connected by a relation $r \in \mathcal{R}$ at timestamp $t \in \mathcal{T}$.

Temporal Knowledge Graph Extrapolation aims to predict a missing object (or subject) entity for a given temporal query, such as $(s, r, ?, t)$ or $(?, r, o, t)$, based on the sequence of historical TKG snapshots $\{\mathcal{G}_{1}, \mathcal{G}_{2}, \dots, \mathcal{G}_{t-1}\}$.
All tasks are formulated as predicting the most plausible object  $o$ from the candidate set $\mathcal{E}$, with inverse relations included.

\section{Our Approach}
As illustrated in Figure~\ref{fig2}, STK-Adapter is integrated into each LLM layer to model evolving structural representations and event chains of TKGs, enabling the LLM to jointly leverage spatial-temporal signals and temporal semantic dependencies for TKG extrapolation.
STK-Adapter involves three MoE modules: ST-MoE, EA-MoE, and CMA-MoE.
The ST-MoE captures spatial structures and temporal patterns embedded in TKGs.
The EA-MoE models intricate temporal semantic dependencies within event chains.
The CMA-MoE facilitates deep cross-modality alignment between the evolving structural representations of TKGs and the semantic space of LLMs.
Finally, an Adaptive Fusion module integrates the outputs of these modules.

\subsection{Spatial-Temporal Mixture of Experts}
ST-MoE aims to leverage specialized experts for layer-wise spatial-temporal modeling, mitigating the progressive dilution of evolving structural information during deep LLM propagation.


\subsubsection{Evolving Graph Encoder}
To enable LLMs to comprehend the evolving structure of TKGs, we adopt a pre-trained Evolving Graph Encoder (e.g., LogCL~\cite{LogCL}) that learns expressive representations of entity and relation evolution in TKGs.
The pre-training process introduces no additional computational overhead during the STK-Adapter fine-tuning.

Given a query $q=(s,r,?,t)$, we construct a sequence of historical subgraphs $\mathcal{G}_{<t} = \{\mathcal{G}_{1}, \dots, \mathcal{G}_{t-1}\}$ to provide essential historical context (details in Appendix \ref{apd:subgraph}).
The historical subgraph sequence $\mathcal{G}_{<t}$ is then encoded by the encoder $f_g$ into evolving structural representations:
\begin{equation}
    \text{H}_\mathcal{E}^{(t)}, \text{H}_\mathcal{R}^{(t)} = f_g(\mathcal{G}_{<t}),
\end{equation}
where $\text{H}_\mathcal{E}^{(t)} \in \mathbb{R}^{|\mathcal{E}| \times d_g}$ and $\text{H}_\mathcal{R}^{(t)} \in \mathbb{R}^{|\mathcal{R}| \times d_g}$ denote entity and relation embedding matrices of dimension $d_g$.
We extract the query-specific subject and relation embeddings, 
$\text{H}_s^{(t)}$, $\text{H}_r^{(t)} \in \mathbb{R}^{1 \times d_g}$, concatenating them to form the initial TKG representation $\text{H}_g^0 \in \mathbb{R}^{1 \times 2d_g}$.
Formally, $\text{H}_g^0= [\text{H}_s^{(t)};\text{H}_r^{(t)}]$, where $[;]$ denotes concatenation operation.
This representation serves as the structural input to the first layer of the ST-MoE module.



\subsubsection{MoE for Evolving Graph}
The ST-MoE is designed to capture diverse spatial structures and temporal patterns inherent in TKG's evolving structural representations.
ST-MoE employs a sparsely activated router that dynamically assigns TKG embeddings to specialized experts, each optimized for spatial-temporal modeling.

At the $l$-th layer of the STK-Adapter, the router $f_{\text{ST\_router}}$ computes routing weights based on the TKG's evolving structural representation $\text{H}_{g}^{l-1}$ from the preceding layer and activates the Top-$k$ experts with the highest routing weights:
\begin{equation}
    \text{gate}^l=f_{\text{ST\_router}}(\text{H}_{g}^{l-1}),
    \mathcal{A}^l = \text{Top-}k(\text{gate}^l),  
\end{equation}
where $\text{gate}^l \in \mathbb{R}^{n}$ denotes the normalized routing weights for $n$ experts, and $\mathcal{A}^l \in \mathbb{N}^{k}$ is the indices of $k$ activated experts at $l$-th layer.

To ensure parameter efficiency, each expert employs a bottleneck structure consisting of a down-projection layer $f_\text{down}$, a non-linear activation $f_\text{nl}$, and an up-projection layer $f_\text{up}$, which enables each expert to model complex spatial-temporal patterns within a specialized latent subspace \cite{MoE_in_DL}.
The $i$-th activated expert operation is defined as:
\begin{equation}
    \text{E}^{(i)}(\text{H}_g^{l-1}) = f^{(i)}_\text{up} \left( f^{(i)}_\text{nl} \left( f^{(i)}_\text{down}(\text{H}_g^{l-1}) \right) \right),
\end{equation}
where $i \in \mathcal{A}^l$, $f^{(i)}_\text{down}$ and $f^{(i)}_\text{up}$ are the unique layers for the $i$-th expert $\text{E}^{(i)}$, and $f^{(i)}_\text{nl}$ is ReLU activation. 
The final output of the ST-MoE is an aggregation of the expert outputs, weighted by their routing values:
\begin{equation}
\overline{\text{H}}_g^l = \sum_{i \in \mathcal{A}^l} \text{gate}_i^l \cdot \text{E}^{(i)}(\text{H}_g^{l-1}).
\end{equation}

The representation $\overline{\text{H}}_g^l$ encodes rich spatial-temporal features.
By iteratively capturing TKG's evolving structure, the ST-MoE mitigates information dilution during deep LLM propagation.


\subsection{Event-Aware Mixture of Experts}
EA-MoE aims to activate specialized experts to model the temporal semantic patterns within event chains, guiding the LLM to understand complex event dependencies.


\subsubsection{Temporal Logic Rule Retrieval}
The effectiveness of LLMs for TKG extrapolation depends on the quality of linearized historical contexts for each query \cite{GPT-NeoX}.
Thus, we adopt a Temporal Logic Rule (TLR) retrieval module \cite{TLogic, GenTKG} to extract temporally and semantically relevant historical event chains for the target query.

The TLR retrieval module mines latent temporal logic rules from the TKG and then applies them to retrieve query-specific historical event chains.
These retrieved events are further formatted into instructions, supplying the LLMs with the essential temporal evidence and semantic context for reasoning.
Detailed descriptions of the TLR retrieval process, instruction construction, sensitivity analyses of TLR-retrieved information quality, and analyses of retrieval strategies are provided in Appendix \ref{apd:TLR}.




\subsubsection{MoE for Event Chains}
Events in TKGs frequently exhibit recurrent patterns and intricate temporal dependencies.
Although modeling evolving graph structures effectively captures spatial–temporal features, it remains insufficient for characterizing the temporal dynamics and complex semantic dependencies inherent in event chains \cite{GPT-NeoX}.
To better enable LLMs to learn these complex patterns, we introduce EA-MoE.
While the EA-MoE adopts a similar architecture to ST-MoE, it is specifically designed to model complex temporal ordering, event repetition, and semantic dependencies in event chains.

The core component of EA-MoE is an event-aware router $f_{\text{EA\_router}}$, which assigns a shared routing signal to all tokens belonging to the same event at a given timestamp.
Such a design ensures consistent processing of tokens within the same event and enables effective modeling of complex temporal patterns in event chains.
To implement this, the event chain $\mathcal{C}$ is tokenized and then encoded by self-attention, yielding the intermediate representation $\hat{\text{H}}_{\text{text}}^{l} \in \mathbb{R}^{\left|\mathcal{C}\right| \times d_t}$ at $l$-th LLM layer, where $d_t$ is the LLM hidden dimension.
For the $j$-th token, the routing signal is computed from the hidden state of its corresponding time token $\tau(j)$:
\begin{equation}
\hat{\text{gate}}_j^{l} = \hat{\text{gate}}_{\tau(j)}^{l} = f_{\text{EA\_router}}(\hat{\text{H}}_{\tau(j)}^{l}),
\end{equation}
where $\tau(j)$ maps token $j$ to the position of its corresponding time token, and $\hat{\text{H}}_{\tau(j)}^{l}$ denotes the hidden state of that time token within $\hat{\text{H}}_{\text{text}}^{l}$.
The subsequent expert processing and aggregation follow the operations in Eqs. (3) and (4) to produce the enhanced event chain representations $\overline{\text{H}}_{\text{text}}^{l}$.
The EA-MoE enables experts to specialize in distinct temporal semantic patterns, ultimately guiding the LLM to learn complex dependencies among events.

\subsection{Cross-Modality Alignment Mixture of Experts}
Due to the inherent modality gap between evolving structural representations of TKGs and the semantic space of LLMs, conventional shallow projection methods are often insufficient to achieve effective cross-modality alignment.
The CMA-MoE introduces TKG-guided attention experts that adaptively augment textual event chain representations with spatial-temporal context at each LLM layer, facilitating deep and progressive alignment.

Similar to the routing strategy of ST-MoE, the router $f_{\text{CMA\_router}}$ at the $l$-th STK-Adapter layer leverages the TKG's evolving structural representation $\text{H}_g^{l-1}$ to dynamically activate the Top-$k$ experts.
This operation allows the LLM to prioritize alignment in spatial and temporal perspectives.
Each expert implements a TKG-guided attention module with a parameter-efficient bottleneck design.
Specifically, the event chain representation $\hat{\text{H}}_{\text{text}}^{l}$ serves as the Query $Q$, while the TKG representation $\text{H}_g^{l-1}$  functions as Key $K$ and Value $V$, allowing the expert to enrich the textual representation with the evolving structural knowledge from TKG.
The computation for the $i$-th expert is defined as:
\begin{equation}
Q = \hat{\text{H}}_{\text{text}}^{l} W_Q^{(i)}, K = \text{H}_g^{l-1} W_K^{(i)}, V = \text{H}_g^{l-1} W_V^{(i)},
\end{equation}
\begin{equation}
\overline{\text{H}}_{\text{CMA}, i}^l = \left( \text{Softmax} \left( \frac{Q K^\top}{\sqrt{d_k}} \right) V \right) W_O^{(i)},
\end{equation}
where $W_Q^{(i)} \in \mathbb{R}^{d_t \times d_k}$, $W_K^{(i)}$, $W_V^{(i)} \in \mathbb{R}^{2d_g \times d_k}$ are expert-specific linear projection matrices to a bottleneck dimension $d_k$, and $W_O^{(i)} \in \mathbb{R}^{d_k \times d_t}$ is an up-projection back to the LLM hidden dimension.
The outputs $\overline{\text{H}}_{\text{CMA},i}^l \in \mathbb{R}^{\left|\mathcal{C}\right| \times d_t}$ from activated experts are aggregated via routing weights to produce the structurally enhanced text representations $\overline{\text{H}}_{\text{CMA}}^l$.
By iteratively infusing TKG features across layers, the CMA-MoE progressively refines event chains and deeply aligns the LLM's semantic space with the TKG's evolving structural representations.

\subsection{Adaptive Fusion}
The Adaptive Fusion Module aggregates outputs from the ST-MoE, EA-MoE, and CMA-MoE modules using distinct fusion strategies for textual and TKG representations.

For the textual pathway, a dynamic gating network $f_\text{gate}$ integrates the refined text features from the EA-MoE, $\overline{\text{H}}_{\text{text}}^l$, and the CMA-MoE, $\overline{\text{H}}_{\text{CMA}}^l$.
The process produces an enhanced textual representation $\text{H}_{\text{enhanced}}^l \in \mathbb{R}^{\left|\mathcal{C}\right| \times d_t}$ that captures both temporal semantic dependencies in event chains and evolving structure of TKGs:
\begin{equation}
\text{H}_{\text{enhanced}}^l = f_\text{gate}(\overline{\text{H}}_{\text{CMA}}^l, \overline{\text{H}}_{\text{text}}^l).
\end{equation}

The enhanced representation is then combined with the text hidden state $\hat{\text{H}}_\text{text}^{l}$ via a residual connection and fused with the feed-forward output $\text{H}_\text{FFN}^l \in \mathbb{R}^{\left|\mathcal{C}\right| \times d_t}$, resulting in the final text representation $\text{H}_\text{text}^l$ for the next LLM layer:
\begin{equation}
\text{H}_\text{text}^l = \hat{\text{H}}_\text{text}^{l} + \text{H}_{\text{enhanced}}^l + \text{H}_\text{FFN}^l.
\end{equation}

For the TKG pathway, the ST-MoE output $\overline{\text{H}}_{g}^l$ is combined with the initial embedding $\text{H}_g^0$:
\begin{equation}
\text{H}_g^l = \overline{\text{H}}_{g}^l + \text{H}_g^0,
\end{equation}
where $\text{H}_g^0$ acts as a structural reference to alleviate information loss.

\subsection{Training and Inference}
During training, the final textual representations are fed into the LLM head to autoregressively predict the target entity.
The model is optimized with a composite loss consisting of cross-entropy and a load-balancing loss for expert utilization.
\begin{equation}
\mathcal{L} = - \sum_{i=1}^{|Y|} \log P(y_i | y_{<i}, \mathcal{C}, \mathcal{G}) + \alpha \sum_{j=1}^{n} f_j \cdot p_j,
\end{equation}
where $Y = \{y_1, \dots, y_{|Y|}\}$ denotes the target entity token sequence, $\mathcal{C}$ represents input event chains, $n$ is the number of experts, $\alpha$ is the balance coefficient, $f_j$ and $p_j$ represent the routing fraction and average routing weights of expert $j$, respectively.

During inference, beam search is used to generate candidate entity sequences, retaining the Top-$B$ most probable outputs at the decoding step \cite{lama-2-7b-CoH}. 
The specific implementation details are described in Appendix \ref{apd:decode}.
To improve ranking accuracy, we adopt a hybrid scoring strategy \cite{LLM-DA} that combines the LLM decoding score $S_{\text{LLM}}(o)$ with a topology-aware score $S_{\text{TKG}}(o)$:
\begin{equation}
S(o) = (1 - \lambda) \cdot S_{\text{LLM}}(o) + \lambda \cdot S_{\text{TKG}}(o),
\end{equation}
where $\lambda \in [0,1]$ is a weighting coefficient, $S_{\text{TKG}}(o)$ is the normalized score obtained by applying a scoring function to the representation by $f_g$.

\section{Experiments}
\subsection{Experimental Setting}
\subsubsection{Datasets}
We evaluate STK-Adapter on four TKG datasets: ICE14, ICE18, ICE15 \cite{ICEWS}, and WIKI \cite{Wiki}. 
Following the chronological splitting in \citet{RE-GCN}, we use an 8:1:1 ratio for training, validation, and testing.
Detailed statistics are provided in Appendix \ref{apd:dataset}.

\begin{table*}[ht]
\caption{Overall performance comparison. The best results are in boldface, and the second-best results are underlined.}
\label{tab:performance}
\centering
\setlength\tabcolsep{4.5pt}
\scriptsize
\begin{tabular}{@{}l|ccc|ccc|ccc|ccc@{}}
\toprule
\multirow{2}{*}{Model} & 
\multicolumn{3}{c@{}}{\textbf{ICE14}} & 
\multicolumn{3}{c@{}}{\textbf{ICE18}} & 
\multicolumn{3}{c@{}}{\textbf{ICE15}} &
\multicolumn{3}{c@{}}{\textbf{WIKI}} 
\\
\cmidrule(lr){2-4} \cmidrule(lr){5-7} \cmidrule(lr){8-10} \cmidrule(lr){11-13}
 & Hit@1 & Hit@3 & Hit@10  & Hit@1 & Hit@3 & Hit@10  & Hit@1 & Hit@3 & Hit@10 & Hit@1 & Hit@3 & Hit@10\\ 
\midrule
REGCN & 30.66 & 44.96 & 59.21 & 21.01 & 34.34 & 48.75 & 37.33 & 53.85 & 68.27 & 73.75 & 80.38 & 83.68 \\
TiRGN & 33.83 & 48.95 & 63.84 & 23.19 & 37.09 & 54.22 & 39.25 & 56.13 & 70.71 & 77.77 & 85.12 & 87.08 \\
CognTKE & 36.49 & 51.11 & 64.49 & 25.21 & 39.93 & 54.71 & 42.62 & 59.42 & 72.70 & 80.01 & 86.07 & 87.34\\
LogCL & 37.76 & 54.71 & \underline{70.26} & 24.53 & 40.32 & \underline{57.74} & \underline{46.07} & \underline{63.72} & \underline{77.87} & 70.85 & 79.11 & 85.59 \\
\midrule
GPT-NeoX & 32.39 & 45.03 & 55.75 & 19.17 & 30.34 & 40.89 & 35.70 & 49.81 & 58.62 & 54.34 & 61.98 & 64.80 \\
CoH & 34.63 & 47.92 & 59.36 & 20.77 & 36.17 & 52.31 & 36.39 & 51.83 & 66.35 & 74.00 & 79.54 & 85.43 \\
GenTKG & 34.87 & 47.34 & 61.91 & 21.50 & 36.62 & 49.57 & 36.01 & 52.54 & 68.67 & 72.81 & 80.24 & 84.60 \\
MESH & 35.22 & 50.42 & 64.25 & 23.61 & 38.46  & 54.18 & 38.62 & 54.95 & 68.60 & 75.03 & 82.98 & 86.37 \\
LLM-DA (REGCN) & 35.13 & 50.99 & 65.16 & 22.20 & 35.92 & 51.35 & 38.68 & 55.57 & 69.63 & 77.82 & 81.53 & 83.12 \\
LLM-DA (TiRGN) & 36.82 & 53.19 & 67.57 & 22.53 & 36.29  & 51.81 & 39.22 & 55.93 & 69.98 & 79.73 & 84.43 & 85.45 \\
LLM-DA (CognTKE) & 36.83 & 52.30 & 65.58 & 22.94 & 37.07 & 52.91 & 39.51 & 57.19 & 71.91 & \underline{84.18} & 85.68 & 86.71 \\
LLM-DA (LogCL) & 37.71 & 53.14 & 66.62 & 22.83 &  36.63 & 52.08 & 40.90 & 56.93 & 70.70 & 79.10 & 83.75 & 84.47 \\
\midrule
STK-Adapter (REGCN) & 37.26 & 52.86 & 67.57 & 23.38 & 39.53 & 54.28 & 42.33 & 55.31 & 69.62 & 82.14 & 87.08 & 88.43 \\
STK-Adapter (TiRGN) & 38.51 & 53.73 & 67.78 & 25.60 & 43.71 & 55.49 & 44.37 & 61.72 & 72.68 & 81.54 & \underline{87.15} & \underline{88.44} \\
STK-Adapter (CognTKE) & \underline{40.20} & \underline{54.91} & 68.09 & \textbf{26.88} & \textbf{45.91} & 57.56 & 44.62 & 62.83 & 74.11 & \textbf{84.38} & 86.98 & 87.88 \\
STK-Adapter (LogCL) & \textbf{41.16} & \textbf{59.03} & \textbf{70.73} & \underline{25.95} & \underline{44.68} & \textbf{59.42} & \textbf{48.82} & \textbf{65.83} & \textbf{78.22} & 82.43 & \textbf{87.22} & \textbf{88.53} \\
\bottomrule
\end{tabular}
\end{table*}

\subsubsection{Baselines}
To validate the effectiveness of STK-Adapter, we compare it with 9 up-to-date methods, including REGCN \cite{RE-GCN}, TiRGN \cite{TiRGN}, LogCL \cite{LogCL}, CognTKE \cite{CognTKE}, GPT-NeoX \cite{GPT-NeoX}, GenTKG \cite{GenTKG}, CoH \cite{lama-2-7b-CoH}, LLM-DA \cite{LLM-DA}, and MESH \cite{MESH}.
Descriptions of these baselines are provided in Appendix \ref{apd:baseline}.

\subsubsection{Implementation Details}
We adopt Llama3-8B~\cite{llama3} as the frozen backbone LLM with a hidden dimension of $d_t {=} 4096$ and only fine-tune the STK-Adapter for 2 epochs.
The evolving graph encoder is pre-trained with $d_g {=} 200$ and kept frozen during fine-tuning.
STK-Adapter employs three MoE modules with $n=4$ experts each, Top-$1$ routing, and an internal dimension of $d_k{=}8$.
The beam search uses $B{=}20$.
All baseline hyperparameters follow the original papers.
More details are provided in Appendix \ref{apd:parameter}.

\subsubsection{Metrics}
We evaluate performance using the Hit@$K$ metric with $K\in\{1,3,10\}$, which measures the proportion of instances in which the ground-truth entity ranks among the Top-$K$ candidates.
Since LLMs cannot explicitly score all entities, we employ beam search \cite{beamsearch} over the LLM's output probability to generate the high-confidence candidate sequences. 
The Hit@$1/3/10$ is then computed by checking whether the ground-truth entity appears in these sequences and recording its rank \cite{lama-2-7b-CoH}.

\subsection{Experimental Results}
\subsubsection{Overall Performance}
Table \ref{tab:performance} presents the results of STK-Adapter and all baselines. 
The STK-Adapter consistently achieves the best performance across all datasets.
While MESH and LLM-DA, which integrate TKG representations with LLMs, outperform single-modality approaches on ICE14 and WIKI, their advantage diminishes on the more complex ICE15 and ICE18 datasets.
The decline is primarily due to their shallow integration strategies, which fail to capture the intricate event dependencies in complex datasets. 
Specifically, MESH aligns textual and TKG embeddings via an MLP and combines them with weighted aggregation, while LLM-DA incorporates the TKG score only at inference without internalizing evolving structure into the LLM.
STK-Adapter employs a layer-wise integration strategy, capturing TKG representations at each LLM layer and deeply aligning them with the LLM's semantic space.

\begin{figure}[ht]
\centering
    \begin{minipage}[t]{1\linewidth}
    \resizebox{1\linewidth}{!}{
        \centering
        \includegraphics[width=1\linewidth]{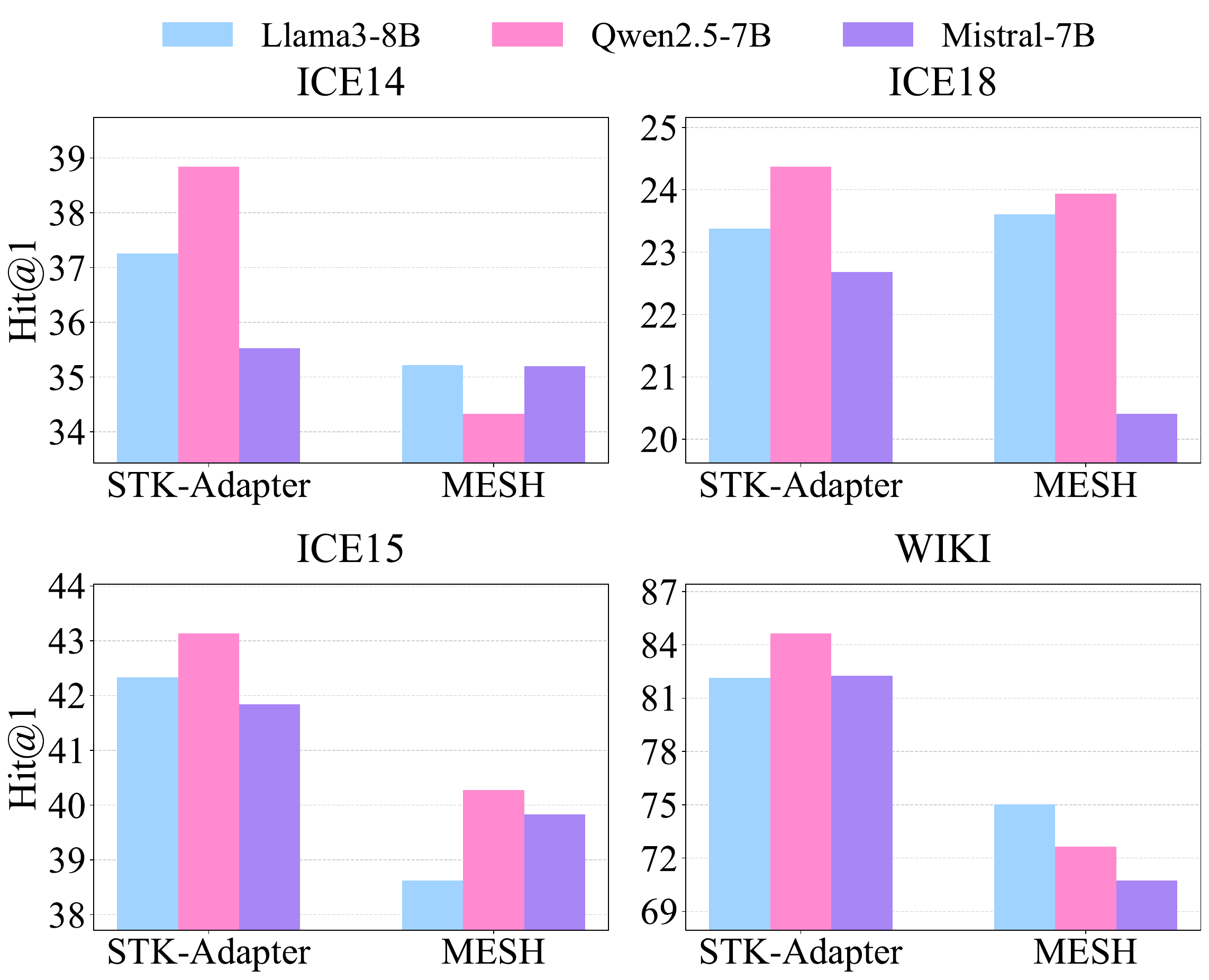}
        }
    \end{minipage}
\caption{A comparison of Hit@1 performance between STK-Adapter and MESH on four datasets, using three backbone LLMs: Llama3-8B, Qwen2.5-7B, and Mistral-7B.}
\label{tab:compatibility}
\end{figure}

\subsubsection{Compatibility Analysis}
To assess the compatibility of STK-Adapter with different evolving graph encoders, we evaluate its performance with four pre-trained encoders: REGCN, TiRGN, CognTKE, and LogCL.
As shown in Table \ref{tab:performance}, STK-Adapter significantly outperforms the best baseline, LLM-DA, when using the same encoder.
The advantage stems from the ST-MoE module, which serves as an encoder-agnostic processor capable of effectively extracting essential features from the output of any given encoder.
Then, the CMA-MoE module deeply aligns these features with the LLM's semantic space, ensuring consistent comprehension regardless of the feature source.
Notably, the result underscores that STK-Adapter works with any pre-trained evolving graph encoder without requiring end-to-end retraining, reducing the practical implementation costs.

We also evaluate STK-Adapter across different LLMs by comparing it with MESH, a near SOTA method that integrates TKG with LLMs, using three mainstream LLMs: Llama3-8B, Qwen2.5-7B, and Mistral-7B.
For a fair comparison, both models employ REGCN as the evolving graph encoder.
Figure \ref{tab:compatibility} shows that STK-Adapter integrates seamlessly with all tested LLMs, while maintaining a consistent performance advantage over MESH.
These results indicate that, regardless of the LLM used, STK-Adapter's deep, layer-wise integration strategy enables effective utilization of TKG knowledge, leading to stable performance improvements.

\begin{table}[ht]
\caption{The results of ablation studies.}
\label{tab:ablation studies}
\centering
\setlength\tabcolsep{1pt}
\scriptsize
\begin{tabular}{lcccccc}
    \toprule
    & \multicolumn{3}{c}{\textbf{ICE14}} & \multicolumn{3}{c}{\textbf{WIKI}} \\
    \cmidrule(lr){2-4} \cmidrule(lr){5-7}
    \textbf{Method} & \textbf{Hit@1} & \textbf{Hit@3} & \textbf{Hit@10} & \textbf{Hit@1} & \textbf{Hit@3} & \textbf{Hit@10} \\
    \midrule
    \textit{STK-Adapter} & \textbf{37.26} & \textbf{52.86} & \textbf{67.57} & \textbf{82.14} & \textbf{87.08} & \textbf{88.43} \\
    \textit{-w/o EA-MoE} & 35.56 & 49.38 & 59.59 & 74.36 & 80.00 & 81.61 \\
    \textit{-w/o ST-MoE} & 37.03 & 52.06 & 63.42 & 80.11 & 85.71 & 87.44 \\
    \textit{-w/o CMA-MoE}& 36.86 & 51.57 & 64.67 & 80.08 & 85.72 & 87.02 \\
    \textit{-w/o ST- \& CMA-MoE} & 33.12 & 48.82 & 63.38 & 78.37 & 83.45 & 84.25 \\
    \textit{-w/o MoE} & 35.48 & 49.92 & 61.49 & 79.78 & 84.74 & 85.49 \\
    \textit{-w/o Hybrid Score} & 37.05 & 52.45 & 65.15 & 81.09 & 86.33 & 87.50 \\
    \textit{-w LoRA} & 32.47 & 45.30 & 57.13 & 72.10 & 75.79 & 76.20 \\
    \bottomrule
\end{tabular}
\end{table}

\subsection{Ablation Studies}
To evaluate the contribution of each STK-Adapter component, we conduct ablation studies on the ICE14 and WIKI datasets, utilizing REGCN as the evolving graph encoder. 
The results are presented in Table \ref{tab:ablation studies}.
The \textit{-w/o EA-MoE} variant removes the Event-Aware MoE module, the \textit{-w/o ST-MoE} variant excludes the Spatial-Temporal MoE module, and the \textit{-w/o CMA-MoE} variant eliminates the Cross-Modality Alignment MoE module, simplifying the fusion to a direct addition of text and TKG representations.
The \textit{-w/o ST- \& CMA-MoE} variant removes the entire TKG processing branch, while \textit{-w/o MoE} replaces the entire MoE architecture with a single standard adapter.
The \textit{-w/o Hybrid Score} variant disables the hybrid scoring strategy during inference, and the \textit{-w LoRA} variant substitutes the STK-Adapter with a LoRA module.

The results show that the variants \textit{-w/o EA-MoE}, \textit{-w/o ST-MoE}, and \textit{-w/o CMA-MoE} consistently perform worse than STK-Adapter, highlighting the effectiveness of each MoE module.
The significant drop in the \textit{-w/o ST- \& CMA-MoE} variant underscores the necessity of integrating TKG's evolving structure into the LLM for reasoning.
Furthermore, the inferior performance of the \textit{-w/o MoE} variant confirms that the MoE design captures more comprehensive and specialized features than a single adapter.
The minimal performance change in the \textit{-w/o Hybrid Score} variant demonstrates that STK-Adapter's strength lies in deep representation alignment and fusion, rather than a dependency on the evolving graph model's raw output scores.
Finally, the results of the \textit{-w LoRA} variant highlight the superiority of STK-Adapter over standard LoRA.

\begin{table}[ht]
\centering
\caption{A comparative analysis of computational efficiency. \textit{Params} are the total model parameters, \textit{Train Params} are the trainable parameters, \textit{Act.Params} are the parameters activated during inference, and \textit{Time} is the inference time per instance.}
\label{tab:efficiency}
\setlength\tabcolsep{2.1 pt} 
\scriptsize
\begin{tabular}{cccccc}
\toprule
\textbf{Model} & \textbf{Params} & \textbf{Train Params} & \textbf{Act. Params} & \textbf{Time} & \textbf{Hit@1}\\
\midrule
CoH & 8,051.23M & 20.97M & 440.08M & 3.87s & 34.63 \\
GenTKG & 8,037.08M & 6.82M & 396.17M & 3.75s & 34.87 \\
MESH & 8,057.28M & 27.02M & 97.54M & 3.71s & 35.22 \\
STK-Adapter & 8,045.53M & 15.27M & 323.77M & 3.63s & 37.26 \\
\bottomrule
\end{tabular}
\end{table}

\subsection{Efficiency Analysis} 
To assess the efficiency of STK-Adapter, we compare it with several Llama3-8B-based methods on the ICE14 dataset.
As shown in Table \ref{tab:efficiency},  STK-Adapter achieves the fastest inference time per instance while maintaining a competitive model size and a relatively small number of trainable parameters.
Compared with LoRA-based methods, including CoH and GenTKG, STK-Adapter achieves superior performance while activating fewer parameters during inference.
Notably, despite incorporating three MoE modules, STK-Adapter maintains high inference efficiency by activating only a small subset of experts per input.
Although MESH activates the fewest parameters, the efficiency stems from its reliance on pre-generated text representations, which largely bypass LLM computation during inference.
Overall, the results demonstrate that STK-Adapter achieves an effective balance between computational efficiency and model performance.

\begin{figure}[ht]
\centering
\begin{minipage}[t]{1\linewidth}
\resizebox{1\linewidth}{!}{
    \centering
    \includegraphics[width=1\linewidth]{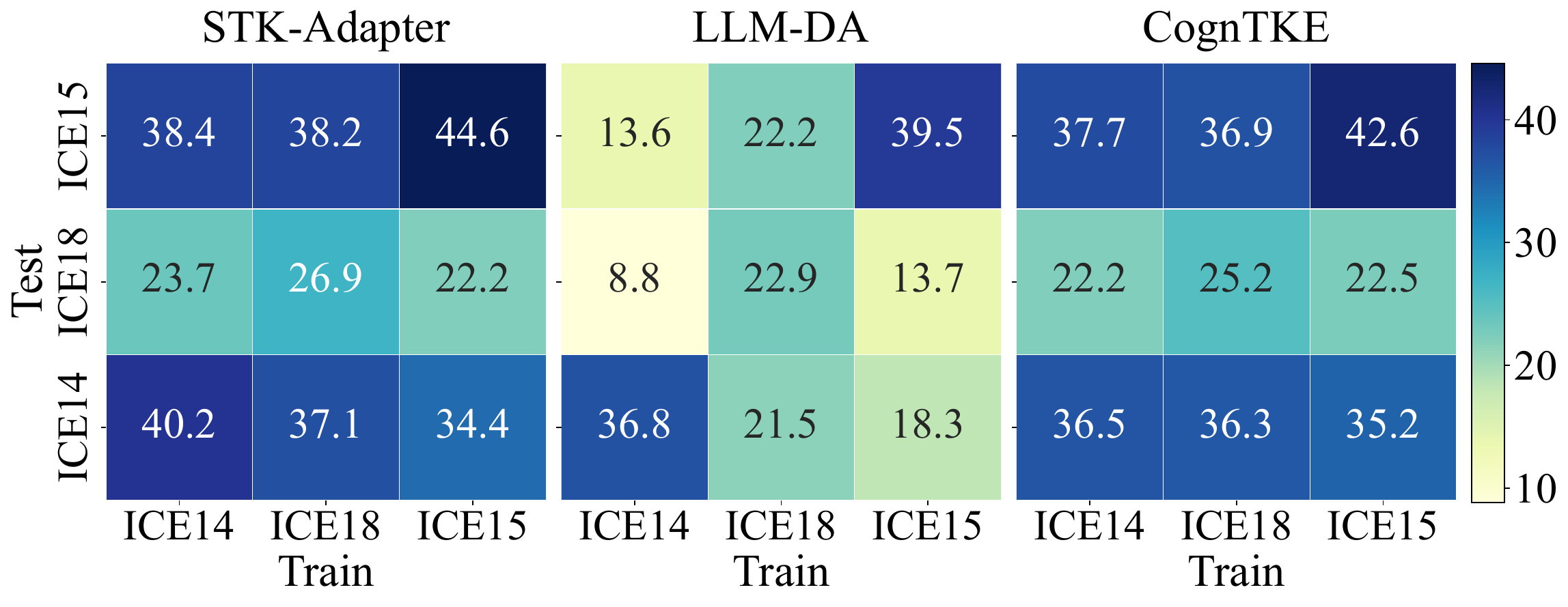}
    }
\end{minipage}
\caption{A comparison of Hit@1 performance among STK-Adapter, LLM-DA, and CognTKE in a cross-dataset generalization setting.}
\label{tab:crossdata}
\end{figure}

\subsection{Cross-Dataset Generalization}
To evaluate the cross-dataset generalization of STK-Adapter, we conduct zero-shot experiments on the ICE series datasets.
Following \citet{CognTKE}, models are trained on one dataset and tested on another without fine-tuning. 
We compare STK-Adapter with the state-of-the-art LLM-DA, employing CognTKE as the evolving graph model for both methods due to its strong inductive reasoning capabilities.
As shown in Figure \ref{tab:crossdata}, STK-Adapter significantly outperforms LLM-DA, which exhibits poor generalization.
The limitation arises from LLM-DA's rule-updating mechanism, which overfits the source distribution and hinders knowledge transfer.
In contrast, STK-Adapter demonstrates generalization, consistently matching or surpassing the inductive baseline CognTKE.
Moreover, STK-Adapter exhibits only a marginal performance drop when transferring across datasets compared to in-dataset evaluation, highlighting its strong knowledge transfer capability across different data distributions.

\begin{figure}[ht]
\centering
\begin{minipage}[t]{1\linewidth}
\resizebox{1\linewidth}{!}{
    \centering
    \includegraphics[width=1\linewidth]{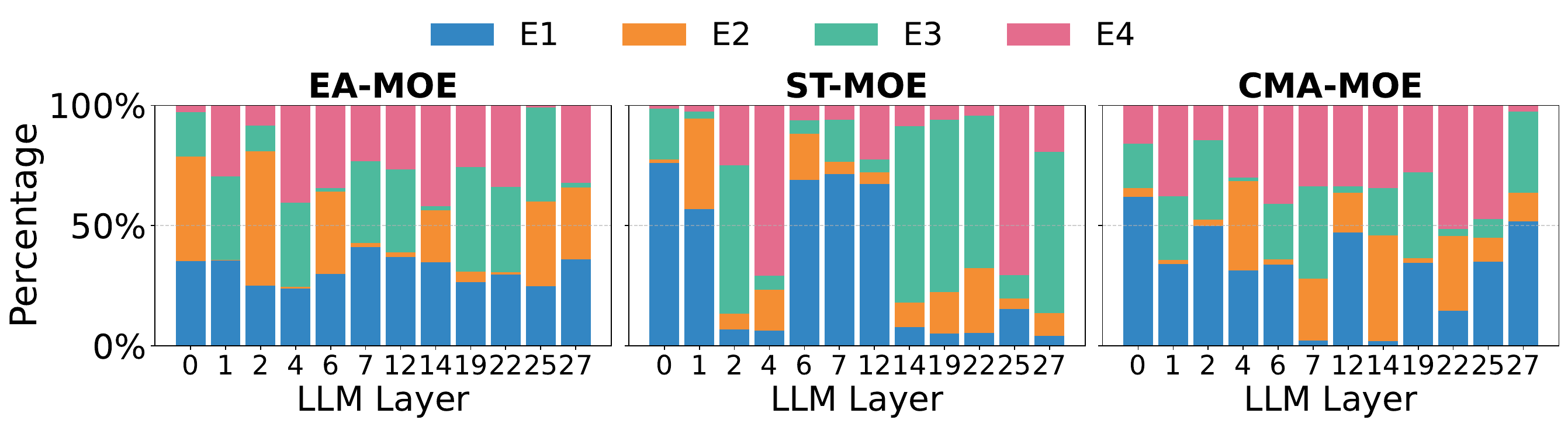}
    }
\end{minipage}
\caption{Distribution of expert routing decisions across the three MoE modules.}
\label{fig:moe}
\end{figure}

\subsection{Expert Routing in MoE Modules}
To analyze the expert routing decisions within our proposed MoE modules and uncover their underlying functional specialization, we visualize the expert activation ratios (E1--E4) across different LLM layers for the ST-MoE, EA-MoE, and CMA-MoE on the ICE14 dataset.
As Figure \ref{fig:moe} illustrates, the result reveals distinct and layer-dependent roles for the experts within each module.

The ST-MoE exhibits a clear hierarchical routing pattern.
Specifically, Expert E1 is preferentially activated in the initial layers, highlighting its specialization in capturing fundamental spatial-temporal representations from the TKG embeddings.
Progressing deeper into the network, Experts E3 and E4 assume greater prominence in the middle and later layers, respectively, to iteratively refine and abstract these foundational features.
The EA-MoE displays a highly collaborative activation pattern. 
This balanced utilization suggests that the complexity of modeling temporal semantic patterns from event chains necessitates the synergistic contributions of multiple specialists.
The CMA-MoE's routing decisions demonstrate a sustained and balanced utilization of experts across all layers. 
Working in concert at every processing stage, the experts ensure the continuous and effective fusion of the TKG structure with the LLM's textual representations.

Collectively, these distinct routing patterns reveal an adaptive strategy. 
The framework dynamically allocates specialized experts at various depths, ensuring that distinct types of information are processed by the most suitable expert.


\section{Conclusion}
In this paper, we propose STK-Adapter, which flexibly integrates an evolving graph encoder with LLMs to facilitate TKG extrapolation.
STK-Adapter involves three MoE modules:
the ST-MoE captures spatial-temporal patterns in TKGs, the EA-MoE models intricate temporal semantic dependencies within event chains, and the CMA-MoE facilitates deep progressive alignment.
Extensive experiments show that STK-Adapter outperforms state-of-the-art methods and exhibits strong generalization capabilities in cross-dataset tasks.


\section*{Limitations}
While STK-Adapter achieves state-of-the-art performance in TKG extrapolation, several limitations warrant further investigation.
Despite its parameter-efficient design, the layer-wise MoE architecture and beam search decoding incur additional memory consumption and computational overhead compared to conventional embedding-based TKG models.


\section*{Acknowledgments}

This work was supported by the Beijing Natural Science Foundation (No.L252034) and Guangxi Natural Science Foundation (Grant No.2026GXNSFBA00640345).

\bibliography{acl_latex}

\begin{thebibliography}{38}
\providecommand{\natexlab}[1]{#1}

\bibitem[{AI@Meta(2024)}]{llama3}
AI@Meta. 2024.
\newblock \href {https://github.com/meta-llama/llama3/blob/main/MODEL_CARD.md} {Llama 3 model card}.

\bibitem[{Boschee et~al.(2015)Boschee, Lautenschlager, O'Brien, Shellman, Starz, and Ward}]{ICEWS}
Elizabeth Boschee, Jennifer Lautenschlager, Sean O'Brien, Steve Shellman, James Starz, and Michael Ward. 2015.
\newblock \href {https://doi.org/10.7910/DVN/28075} {{ICEWS Coded Event Data}}.

\bibitem[{Chang et~al.(2025)Chang, Wu, Tao, Ma, Huang, and Chua}]{TGL-LLM}
He~Chang, Jie Wu, Zhulin Tao, Yunshan Ma, Xianglin Huang, and Tat-Seng Chua. 2025.
\newblock Integrate temporal graph learning into llm-based temporal knowledge graph model.
\newblock \emph{arXiv preprint arXiv:2501.11911}.

\bibitem[{Chen(2024)}]{wan_3}
Huajun Chen. 2024.
\newblock \href {https://doi.org/10.3724/2096-7004.di.2024.0001} {Large knowledge model: Perspectives and challenges}.
\newblock \emph{DATA INTELLIGENCE}, 6(3):587--620.

\bibitem[{Chen et~al.(2025{\natexlab{a}})Chen, Huang, Zhang, Wang, Lin, Chang, and Wan}]{chenwei_2}
Wei Chen, Haoyu Huang, Zhiyu Zhang, Tianyi Wang, Youfang Lin, Liang Chang, and Huaiyu Wan. 2025{\natexlab{a}}.
\newblock Next-poi recommendation via spatial-temporal knowledge graph contrastive learning and trajectory prompt.
\newblock \emph{IEEE Transactions on Knowledge and Data Engineering}.

\bibitem[{Chen et~al.(2024)Chen, Wan, Wu, Zhao, Cheng, Li, and Lin}]{LogCL}
Wei Chen, Huaiyu Wan, Yuting Wu, Shuyuan Zhao, Jiayaqi Cheng, Yuxin Li, and Youfang Lin. 2024.
\newblock Local-global history-aware contrastive learning for temporal knowledge graph reasoning.
\newblock In \emph{2024 IEEE 40th International Conference on Data Engineering (ICDE)}, pages 733--746. IEEE.

\bibitem[{Chen et~al.(2025{\natexlab{b}})Chen, Wu, Guo, Wu, Jiang, Lin, and Wan}]{chenwei_1}
Wei Chen, Yuting Wu, Shengnan Guo, Shuhan Wu, Zhishu Jiang, Youfang Lin, and Huaiyu Wan. 2025{\natexlab{b}}.
\newblock Dual-view temporal knowledge graph reasoning.
\newblock \emph{Knowledge-Based Systems}, page 114330.

\bibitem[{Chen et~al.(2025{\natexlab{c}})Chen, Wu, Wu, Zhang, Liao, Lin, and Wan}]{CognTKE}
Wei Chen, Yuting Wu, Shuhan Wu, Zhiyu Zhang, Mengqi Liao, Youfang Lin, and Huaiyu Wan. 2025{\natexlab{c}}.
\newblock Cogntke: A cognitive temporal knowledge extrapolation framework.
\newblock In \emph{Proceedings of the AAAI Conference on Artificial Intelligence}, volume~39, pages 14815--14823.

\bibitem[{Chen et~al.(2022)Chen, Deng, Wu, Gu, and Li}]{MoE_in_DL}
Zixiang Chen, Yihe Deng, Yue Wu, Quanquan Gu, and Yuanzhi Li. 2022.
\newblock Towards understanding mixture of experts in deep learning.
\newblock \emph{arXiv preprint arXiv:2208.02813}.

\bibitem[{Dasgupta et~al.(2018)Dasgupta, Ray, and Talukdar}]{Wiki}
Shib~Sankar Dasgupta, Swayambhu~Nath Ray, and Partha Talukdar. 2018.
\newblock Hyte: Hyperplane-based temporally aware knowledge graph embedding.
\newblock In \emph{Proceedings of the 2018 conference on empirical methods in natural language processing}, pages 2001--2011.

\bibitem[{Deng et~al.(2025)Deng, Wu, Wang, Zhao, Zhu, Liu, Xu, Fu, Wu, Zheng, Zhao, and Qian}]{MESH}
Yimin Deng, Yuxia Wu, Yejing Wang, Guoshuai Zhao, Li~Zhu, Qidong Liu, Derong Xu, Zichuan Fu, Xian Wu, Yefeng Zheng, Xiangyu Zhao, and Xueming Qian. 2025.
\newblock \href {https://aclanthology.org/2025.findings-acl.1056/} {A multi-expert structural-semantic hybrid framework for unveiling historical patterns in temporal knowledge graphs}.
\newblock In \emph{Findings of the Association for Computational Linguistics, {ACL} 2025, Vienna, Austria, July 27 - August 1, 2025}, Findings of {ACL}, pages 20553--20565. Association for Computational Linguistics.

\bibitem[{Dou et~al.(2024)Dou, Zhou, Liu, Gao, Shen, Xiong, Zhou, Wang, Xi, Fan, Pu, Zhu, Zheng, Gui, Zhang, and Huang}]{Loramoe}
Shihan Dou, Enyu Zhou, Yan Liu, Songyang Gao, Wei Shen, Limao Xiong, Yuhao Zhou, Xiao Wang, Zhiheng Xi, Xiaoran Fan, Shiliang Pu, Jiang Zhu, Rui Zheng, Tao Gui, Qi~Zhang, and Xuanjing Huang. 2024.
\newblock \href {https://doi.org/10.18653/V1/2024.ACL-LONG.106} {Loramoe: Alleviating world knowledge forgetting in large language models via moe-style plugin}.
\newblock In \emph{Proceedings of the 62nd Annual Meeting of the Association for Computational Linguistics (Volume 1: Long Papers), {ACL} 2024, Bangkok, Thailand, August 11-16, 2024}, pages 1932--1945. Association for Computational Linguistics.

\bibitem[{Gao et~al.(2024)Gao, Qiao, Kan, Wen, He, and Li}]{GenTKGQA}
Yifu Gao, Linbo Qiao, Zhigang Kan, Zhihua Wen, Yongquan He, and Dongsheng Li. 2024.
\newblock Two-stage generative question answering on temporal knowledge graph using large language models.
\newblock \emph{arXiv preprint arXiv:2402.16568}.

\bibitem[{Han et~al.(2024)Han, Nguyen, Harris, Ho, and Saria}]{fusemoe}
Xing Han, Huy Nguyen, Carl Harris, Nhat Ho, and Suchi Saria. 2024.
\newblock Fusemoe: Mixture-of-experts transformers for fleximodal fusion.
\newblock \emph{Advances in Neural Information Processing Systems}, 37:67850--67900.

\bibitem[{Houlsby et~al.(2019)Houlsby, Giurgiu, Jastrzebski, Morrone, De~Laroussilhe, Gesmundo, Attariyan, and Gelly}]{Adapter}
Neil Houlsby, Andrei Giurgiu, Stanislaw Jastrzebski, Bruna Morrone, Quentin De~Laroussilhe, Andrea Gesmundo, Mona Attariyan, and Sylvain Gelly. 2019.
\newblock Parameter-efficient transfer learning for nlp.
\newblock In \emph{International conference on machine learning}, pages 2790--2799. PMLR.

\bibitem[{Hu et~al.(2022)Hu, Shen, Wallis, Allen{-}Zhu, Li, Wang, Wang, and Chen}]{lora}
Edward~J. Hu, Yelong Shen, Phillip Wallis, Zeyuan Allen{-}Zhu, Yuanzhi Li, Shean Wang, Lu~Wang, and Weizhu Chen. 2022.
\newblock \href {https://openreview.net/forum?id=nZeVKeeFYf9} {Lora: Low-rank adaptation of large language models}.
\newblock In \emph{The Tenth International Conference on Learning Representations, {ICLR} 2022, Virtual Event, April 25-29, 2022}. OpenReview.net.

\bibitem[{Huang et~al.(2023)Huang, Zhang, Mei, and Ma}]{LLMuseGNN}
Jin Huang, Xingjian Zhang, Qiaozhu Mei, and Jiaqi Ma. 2023.
\newblock Can llms effectively leverage graph structural information through prompts, and why?
\newblock \emph{arXiv preprint arXiv:2309.16595}.

\bibitem[{Jacobs et~al.(1991)Jacobs, Jordan, Nowlan, and Hinton}]{MoE}
Robert~A. Jacobs, Michael~I. Jordan, Steven~J. Nowlan, and Geoffrey~E. Hinton. 1991.
\newblock \href {https://doi.org/10.1162/neco.1991.3.1.79} {Adaptive mixtures of local experts}.
\newblock \emph{Neural Computation}, 3(1):79--87.

\bibitem[{Jiang et~al.(2023)Jiang, Sablayrolles, Mensch, Bamford, Chaplot, de~Las~Casas, Bressand, Lengyel, Lample, Saulnier, Lavaud, Lachaux, Stock, Scao, Lavril, Wang, Lacroix, and Sayed}]{Mistral}
Albert~Qiaochu Jiang, Alexandre Sablayrolles, Arthur Mensch, Chris Bamford, Devendra~Singh Chaplot, Diego de~Las~Casas, Florian Bressand, Gianna Lengyel, Guillaume Lample, Lucile Saulnier, L{\'e}lio~Renard Lavaud, Marie-Anne Lachaux, Pierre Stock, Teven~Le Scao, Thibaut Lavril, Thomas Wang, Timoth{\'e}e Lacroix, and William~El Sayed. 2023.
\newblock \href {https://api.semanticscholar.org/CorpusID:263830494} {Mistral 7b}.
\newblock \emph{ArXiv}, abs/2310.06825.

\bibitem[{Lee et~al.(2023)Lee, Ahrabian, Jin, Morstatter, and Pujara}]{GPT-NeoX}
Dong-Ho Lee, Kian Ahrabian, Woojeong Jin, Fred Morstatter, and Jay Pujara. 2023.
\newblock Temporal knowledge graph forecasting without knowledge using in-context learning.
\newblock \emph{arXiv preprint arXiv:2305.10613}.

\bibitem[{Li et~al.(2022)Li, Sun, and Zhao}]{TiRGN}
Yujia Li, Shiliang Sun, and Jing Zhao. 2022.
\newblock Tirgn: Time-guided recurrent graph network with local-global historical patterns for temporal knowledge graph reasoning.
\newblock In \emph{IJCAI}, pages 2152--2158.

\bibitem[{Li et~al.(2021)Li, Jin, Li, Guan, Guo, Shen, Wang, and Cheng}]{RE-GCN}
Zixuan Li, Xiaolong Jin, Wei Li, Saiping Guan, Jiafeng Guo, Huawei Shen, Yuanzhuo Wang, and Xueqi Cheng. 2021.
\newblock Temporal knowledge graph reasoning based on evolutional representation learning.
\newblock In \emph{Proceedings of the 44th international ACM SIGIR conference on research and development in information retrieval}, pages 408--417.

\bibitem[{Liao et~al.(2025)Liao, Chen, Shen, Guo, and Wan}]{HMoRA}
Mengqi Liao, Wei Chen, Junfeng Shen, Shengnan Guo, and Huaiyu Wan. 2025.
\newblock Hmora: Making llms more effective with hierarchical mixture of lora experts.
\newblock In \emph{The Thirteenth International Conference on Learning Representations}.

\bibitem[{Liao et~al.(2023)Liao, Jia, Li, Ma, and Tresp}]{GenTKG}
Ruotong Liao, Xu~Jia, Yangzhe Li, Yunpu Ma, and Volker Tresp. 2023.
\newblock Gentkg: Generative forecasting on temporal knowledge graph with large language models.
\newblock \emph{arXiv preprint arXiv:2310.07793}.

\bibitem[{Liu et~al.(2022)Liu, Ma, Hildebrandt, Joblin, and Tresp}]{TLogic}
Yushan Liu, Yunpu Ma, Marcel Hildebrandt, Mitchell Joblin, and Volker Tresp. 2022.
\newblock Tlogic: Temporal logical rules for explainable link forecasting on temporal knowledge graphs.
\newblock In \emph{Proceedings of the AAAI conference on artificial intelligence}, volume~36, pages 4120--4127.

\bibitem[{Luo et~al.(2024)Luo, Gu, Li, Li, Lin, Li, and Yang}]{lama-2-7b-CoH}
Ruilin Luo, Tianle Gu, Haoling Li, Junzhe Li, Zicheng Lin, Jiayi Li, and Yujiu Yang. 2024.
\newblock Chain of history: Learning and forecasting with llms for temporal knowledge graph completion.
\newblock \emph{arXiv preprint arXiv:2401.06072}.

\bibitem[{Shazeer et~al.(2017)Shazeer, Mirhoseini, Maziarz, Davis, Le, Hinton, and Dean}]{MoE_Structure}
Noam Shazeer, Azalia Mirhoseini, Krzysztof Maziarz, Andy Davis, Quoc Le, Geoffrey Hinton, and Jeff Dean. 2017.
\newblock Outrageously large neural networks: The sparsely-gated mixture-of-experts layer.
\newblock \emph{arXiv preprint arXiv:1701.06538}.

\bibitem[{Sutskever et~al.(2014)Sutskever, Vinyals, and Le}]{beamsearch}
Ilya Sutskever, Oriol Vinyals, and Quoc~V Le. 2014.
\newblock Sequence to sequence learning with neural networks.
\newblock \emph{Advances in neural information processing systems}, 27.

\bibitem[{Wang et~al.(2024{\natexlab{a}})Wang, Kai, Luo, Wei, Hu, Liew, Pan, and Yin}]{LLM-DA}
Jiapu Wang, Sun Kai, Linhao Luo, Wei Wei, Yongli Hu, Alan Wee-Chung Liew, Shirui Pan, and Baocai Yin. 2024{\natexlab{a}}.
\newblock Large language models-guided dynamic adaptation for temporal knowledge graph reasoning.
\newblock \emph{Advances in Neural Information Processing Systems}, 37:8384--8410.

\bibitem[{Wang et~al.(2024{\natexlab{b}})Wang, Qi, Chen, Huang, and Wu}]{wan_4}
Keyu Wang, Guilin Qi, Jiaoyan Chen, Yi~Huang, and Tianxing Wu. 2024{\natexlab{b}}.
\newblock \href {https://doi.org/10.3724/2096-7004.di.2024.0088} {Embedding ontologies via incorporating extensional and intensional knowledge}.
\newblock \emph{DATA INTELLIGENCE}, 6(4):1222--1241.

\bibitem[{Wang et~al.(2022)Wang, Mukherjee, Liu, Gao, Awadallah, and Gao}]{Aadamix}
Yaqing Wang, Subhabrata Mukherjee, Xiaodong Liu, Jing Gao, Ahmed~Hassan Awadallah, and Jianfeng Gao. 2022.
\newblock Adamix: Mixture-of-adapter for parameter-efficient tuning of large language models.
\newblock \emph{arXiv preprint arXiv:2205.12410}, 1(2):4.

\bibitem[{Wu et~al.(2026)Wu, Song, Yao, Han, Wan, Lin, and Lv}]{wuhao}
Hao Wu, Shoucheng Song, Chang Yao, Sheng Han, Huaiyu Wan, Youfang Lin, and Kai Lv. 2026.
\newblock Think how your teammates think: Active inference can benefit decentralized execution.
\newblock In \emph{Proceedings of the AAAI Conference on Artificial Intelligence}, volume~40, pages 29749--29757.

\bibitem[{Yang et~al.(2024)Yang, Yang, Hui, Zheng, Yu, Zhou, Li, Li, Liu, Huang, Dong, Wei, Lin, Tang, Wang, Yang, Tu, Zhang, Ma, Xu, Zhou, Bai, He, Lin, Dang, Lu, Chen, Yang, Li, Xue, Ni, Zhang, Wang, Peng, Men, Gao, Lin, Wang, Bai, Tan, Zhu, Li, Liu, Ge, Deng, Zhou, Ren, Zhang, Wei, Ren, Fan, Yao, Zhang, Wan, Chu, Liu, Cui, Zhang, and Fan}]{qwen2.5}
An~Yang, Baosong Yang, Binyuan Hui, Bo~Zheng, Bowen Yu, Chang Zhou, Chengpeng Li, Chengyuan Li, Dayiheng Liu, Fei Huang, Guanting Dong, Haoran Wei, Huan Lin, Jialong Tang, Jialin Wang, Jian Yang, Jianhong Tu, Jianwei Zhang, Jianxin Ma, and 40 others. 2024.
\newblock Qwen2 technical report.
\newblock \emph{arXiv preprint arXiv:2407.10671}.

\bibitem[{Yao et~al.(2025)Yao, Lin, Song, Wu, Ma, Han, and Lv}]{yaochang}
Chang Yao, Youfang Lin, Shoucheng Song, Hao Wu, Yuqing Ma, Shang Han, and Kai Lv. 2025.
\newblock From general relation patterns to task-specific decision-making in continual multi-agent coordination.
\newblock \emph{arXiv preprint arXiv:2507.06004}.

\bibitem[{Yin et~al.(2026)Yin, Wang, Liu, Li, Yao, Li, Li, Ren, and Yang}]{yinzihao}
Zihao Yin, Zhihai Wang, Haiyang Liu, Chuanlan Li, Muyun Yao, Shijiang Li, Fangjing Li, Jia Ren, and Yanchao Yang. 2026.
\newblock Pua: Pseudo-features made useful again for robust graph node classification under distribution shift.
\newblock \emph{Pattern Recognition}, page 113185.

\bibitem[{Zhang et~al.(2025{\natexlab{a}})Zhang, Zhang, Chu, Wu, Li, and Wei}]{More}
Dacao Zhang, Kun Zhang, Shimao Chu, Le~Wu, Xin Li, and Si~Wei. 2025{\natexlab{a}}.
\newblock More: A mixture of low-rank experts for adaptive multi-task learning.
\newblock \emph{arXiv preprint arXiv:2505.22694}.

\bibitem[{Zhang et~al.(2025{\natexlab{b}})Zhang, Chen, Lin, and Wan}]{zhiyu}
Zhiyu Zhang, Wei Chen, Youfang Lin, and Huaiyu Wan. 2025{\natexlab{b}}.
\newblock A generative adaptive replay continual learning model for temporal knowledge graph reasoning.
\newblock \emph{arXiv preprint arXiv:2506.04083}.

\bibitem[{Zhao et~al.(2025)Zhao, Chen, Shi, Zhou, Lin, and Wan}]{zhaoshuyuan}
Shuyuan Zhao, Wei Chen, Boyan Shi, Liyong Zhou, Shuohao Lin, and Huaiyu Wan. 2025.
\newblock Spatial-temporal knowledge distillation for takeaway recommendation.
\newblock In \emph{Proceedings of the AAAI Conference on Artificial Intelligence}, volume~39, pages 13365--13373.

\end{thebibliography}

\appendix

\section{Sequence of Historical Subgraph Sampling}
\label{apd:subgraph}
We employ an efficient neighborhood sampling strategy to extract historical subgraph sequences $\mathcal{G}_{<t}$ from TKGs.

For a given query $(s,r, ?, t)$, we treat each preceding  timestamp $t'\in \{1,2,\dots, {t-1}\}$ as a discrete temporal snapshot and define the source entity $s$ serving as the root node. 
Within each snapshot, we sample a fixed number of neighbors adjacent to $s$, while preserving the specific relations between the central node and its sampled neighbors. 
Furthermore, the same sampling procedure is recursively applied to the sampled neighbors to capture higher-order topological information until a predefined depth $m$ is reached.
Consequently, a temporal snapshot subgraph $\mathcal{G}_{t'}$ is constructed for each timestamp. 
These subgraphs are organized chronologically to form the historical subgraph sequence $\mathcal{G}_{<t} = \{\mathcal{G}_{1}, \mathcal{G}_{2}, \dots, \mathcal{G}_{t-1}\}$.

\section{Historical Event Retrieval Strategy}
\label{apd:TLR}
\subsection{Temporal Logic Rule Process}
\label{apd:TLR_process}
The concept of a Temporal Logic Rule is first introduced by TLogic \cite{TLogic} and later utilized by GenTKG \cite{GenTKG} to convert TKGs into linearized event chains. 
Specifically, a TLR $\rho$ is designed to formally describe the historical conditions on which the existence of a relation $r$ between entities $s$ and $o$ at a specific timestamp $t_q$ depends. 
The general form is defined as:
\begin{equation}
    \rho:=r(s, o, t_q)\leftarrow\land_{i=1}^{q-1}r^*(s, o, t_i),
\end{equation}
where the left-hand side $r(s, o, t_q)$ denotes the rule head, representing the relation $r$ that holds at time $t_q$. 
The right-hand side is the rule body, which consists of a set of historical relations $r^*$ ordered chronologically from $t_1$ to $t_{q-1}$. 
The arrow ($\leftarrow$) signifies that the rule head is inferred from the rule body.

The critical phase of this strategy is rule mining, which aims to discover high-quality TLRs for a specific relation $r$. 
The process commences by defining a known fact $(s, r, ?, t_q)$ as the rule head. Subsequently, to construct the rule body, a time-decreasing random walk is initiated from the object entity $o$. 
This walk adheres to the following exponentially weighted transition distribution:
\begin{equation}
    P(u; o, t_q) = \frac{\exp(t_u - t_q)}{\sum_{\hat{u}\in C(o,t_q)}exp(t_{\hat{u}}-t_q)}.
\end{equation}

This distribution function preferentially samples candidate edges from the neighborhood set $C(o,t_q)$ of entity $o$ whose timestamps $t_u$ are closer to the target time $t_q$, thereby generating paths that constitute the rule body. By computing the confidence of the rules corresponding to these paths, we then filter and retain the high-confidence rules to form the final rule set.

\begin{figure}[htbp]
    \centering
    \includegraphics[width=1 \linewidth]{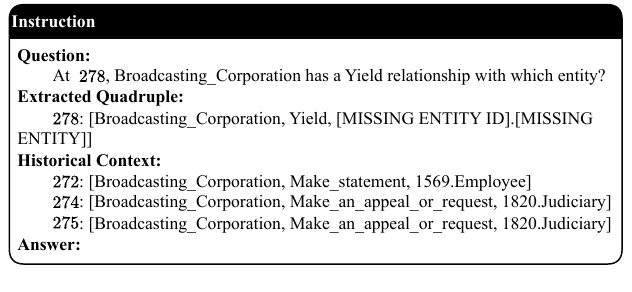}
    \caption{An example of the instruction format.}
    \label{fig: instruction}
\end{figure}

\subsection{Instruction Construction}
Instruction tuning enhances the instruction-following capabilities of LLMs by fine-tuning them on curated prompt-response pairs.
In our proposed STK-Adapter framework, we formulate the TKG extrapolation task as a generative instruction-following task; the specific format of the instruction is illustrated in Figure \ref{fig: instruction}.


Given a query $(s, r, ?, t)$, the historical context is modeled as a sequence of chronologically ordered event chains retrieved during the TLR phase (detailed in Appendix \ref{apd:TLR_process}). 
Each historical fact is linearized and formatted into a predefined template: $t : [s, r, o_{id}.o]$.
Correspondingly, the query is represented in a prefix form, $t : [s, r, $ which guides the LLM to predict the target entity through autoregressive generation.
To ensure a fair comparison with existing baselines, we adopt the indexing strategy established by \citet{GPT-NeoX}, adding a unique numerical index $o_{id}$ before each candidate object $o$.
Evaluation metrics are subsequently computed based on these generated indices.

\subsection{Sensitivity to the Quality of TLR- Retrieved Information}
To evaluate the impact of TLR quality, we randomly remove 50\%, 80\%, and 100\% of the retrieved historical event chains to simulate reduced richness of TLR-retrieved information, while retaining the TKG's evolving structural representations.
For fair comparison, all experiments utilized Llama3-8B as the LLM and REGCN as the evolving graph encoder. 
The results are shown in Table \ref{tab:TLR quality}.
As the proportion of removed events increases, performance declines consistently, indicating that the model effectiveness depends on the richness of the information retrieved by the TLR module.
Furthermore, when both historical event chains and TKG representations are completely removed, the model degrades to a Pure LLM. Notably, even when 100\% of historical event chains are removed, the model still marginally outperforms the Pure LLM, indicating that the model enables LLMs to leverage evolving structural information from the TKG for reasoning. 

\begin{table}[ht]
\centering
\caption{Impact of TLR-retrieved information quality on performance.}
\label{tab:TLR quality}
\setlength\tabcolsep{2pt} 
\scriptsize
\begin{tabular}{lcccccc}
\toprule
& \multicolumn{3}{c}{\textbf{ICE14}} & \multicolumn{3}{c}{\textbf{WIKI}} \\
\cmidrule(lr){2-4} \cmidrule(lr){5-7}
\textbf{Removal Proportion} & \textbf{Hit@1} & \textbf{Hit@3} & \textbf{Hit@10} & \textbf{Hit@1} & \textbf{Hit@3} & \textbf{Hit@10} \\
\midrule
0\% & 37.26 & 52.86 & 67.57 & 82.14 & 87.08 & 88.43 \\
50\% & 32.51 & 46.20 & 59.65 & 64.81 & 81.73 & 85.86 \\
80\% & 27.94 & 38.75 & 56.20 & 55.68 & 68.75 & 71.80 \\
100\% & 18.47 & 25.12 & 50.50 & 37.29 & 55.81 & 64.02 \\
Pure LLM & 13.45 & 19.86 & 44.36 & 27.14 & 45.97 & 53.73 \\
\bottomrule
\end{tabular}
\end{table}

\subsection{Sensitivity to Different Retrieval Strategies}
To assess the sensitivity of different retrieval strategies on STK-Adapter, we compared the performance of the TLR-based approach with a pattern matching strategy \cite{lama-2-7b-CoH} and a hybrid retrieval strategy on the ICE14 and WIKI. 
For comparison, all experiments utilized Llama3-8B as the LLM and REGCN as the evolving graph encoder. 
The results, presented in Table \ref{tab:TLR}, indicate that the TLR strategy outperforms pattern matching, which can be attributed to the higher-quality historical context it provides.
Notably, the performance difference between the three strategies is minimal. 
The result highlights that STK-Adapter exhibits low sensitivity to the choice of retrieval strategy. 
The model's superior performance stems primarily from its ability to deeply comprehend and fuse TKG information with textual semantics, rather than from a dependency on a specific retrieval method.
Although the hybrid strategy incorporates both semantic similarity and logical rules, it yields only limited improvements while introducing additional token overhead.

\begin{table}[ht]
\centering
\caption{Performance comparison of different retrieval strategies.} 
\label{tab:TLR}
\setlength\tabcolsep{4pt} 
\scriptsize 
\begin{tabular}{llccc}
\toprule
\multirow{2}{*}{\textbf{Dataset}} & \multirow{2}{*}{\textbf{Metric}} & \multicolumn{3}{c}{\textbf{Method}} \\
\cmidrule(lr){3-5}
& & \textbf{TLR} & \textbf{Pattern Matching} & \textbf{Hybrid Retrieval} \\
\midrule
\multirow{3}{*}{\textbf{ICE14}} 
& Hit@1  & 37.26 & 36.59 & 37.69 \\
& Hit@3  & 52.86 & 52.34 & 53.24 \\
& Hit@10 & 67.57 & 67.25 & 67.96 \\
& Avg Tokens & 856 & 737 & 1260 \\
\midrule 
\multirow{3}{*}{\textbf{WIKI}}  
& Hit@1  & 82.14 & 81.10 & 82.88 \\
& Hit@3  & 87.08 & 86.33 & 87.24 \\
& Hit@10 & 88.43 & 87.51 & 89.01 \\
& Avg Tokens & 792 & 838 & 953 \\
\bottomrule
\end{tabular}
\end{table}

\section{Decoding Strategy}
\label{apd:decode}
The evaluation methodologies for embedding-based and LLM-based approaches exhibit a fundamental divergence.
Embedding-based methods can compute scores for all entities within the TKG, thereby enabling the generation of a precise global ranking. 
In contrast, the generative architecture of LLMs precludes the direct scoring and ranking of the entire candidate entity set. 
To overcome the constraint, we adhere to the common practice in the field \cite{lama-2-7b-CoH, GenTKG, GPT-NeoX} and employ a beam search strategy. 
The strategy generates a constrained list of high-confidence candidates based on the model's decoding probabilities. Subsequently, we calculate Hit@$1/3/10$ metrics based on the generated list.

Beam search is a decoding strategy that maintains $k$ beams of possible generated responses at each time step $t$.
The generation of response is updated as follows:
\begin{equation}
B_t = \underset{r_{1:t} \in V^t, |B_t|=k}{\text{arg top-k}} \left( \sum_{i=1}^{t} \log P(r_i | r_{1:i-1}) \right).
\end{equation}

For each generated response, the $k$ tokens with the highest probabilities are selected based on these.
The results in $k\times k$ new response candidates.
The next $k$ beams of response are obtained by selecting the Top-$k$ responses with the highest probabilities from the generated response candidates.
The highest probability is determined by the product of probabilities of $|R|$ tokens that constitute the response, where $|R|$ represents the length of the current response.

In this context, the single-step setting is employed, wherein for each test query in the test dataset, the model can access the ground truth from past timestamps. 
Consequently, after the prediction for this step is completed, the ground truth from the current timestamp is added to the history of the next timestamp before its execution.

\section{Dataset Details}
\label{apd:dataset}
To evaluate the performance of STK-Adapter, we selected four benchmark datasets from the field of TKGR: ICE14, ICE18, ICE15 \cite{ICEWS}, and WIKI \cite{Wiki}. The ICE series datasets are derived from the Integrated Crisis Early Warning System \footnote{The datasets are available at: http://www.icews.com/} and consist of political events with precise timestamps. The WIKI dataset is a temporal knowledge graph constructed from Wikipedia. Detailed statistical characteristics are provided in Table \ref{tab:datasets}.

\begin{table}[ht]
\caption{Details of the TKG datasets.}
\label{tab:datasets}
\centering
\scriptsize
\begin{tabular}{c|cccc}
\toprule
\textbf{Dataset} & \textbf{ICE14} & \textbf{ICE18} & \textbf{ICE05-15} & \textbf{WIKI} \\
\midrule
Entities & 7,128 & 23,033 & 19,094 & 12,554 \\
Relations & 230 & 256 & 251 & 24 \\
Training & 74,845 & 373,018 & 368,868 & 539,286 \\
Validation & 8,514 & 45,995 & 46,302 & 67,538 \\
Test & 7,371 & 49,545 & 46,159 & 63,110 \\
Time granularity & 24 hours & 24 hours & 24 hours & 1 year \\
\bottomrule
\end{tabular}
\end{table}

\section{Baseline Descriptions}
\label{apd:baseline}
To comprehensively evaluate the performance of STK-Adapter, we compare it against a diverse set of baseline models. These baselines are categorized into three main groups: (1) embedding-based methods, (2) LLM-based methods, and (3) hybrid approaches that integrate LLMs with TKGs.
Below is a brief overview of each method:

\begin{itemize}
    \item \textbf{REGCN} \cite{RE-GCN} models TKG evolution by using a GCN to capture structure at each timestamp and an RNN to learn the temporal dynamics of entity representations.
    \item \textbf{TiRGN} \cite{TiRGN} captures sequential, repetitive, and cyclical patterns of historical facts through a local-global encoder architecture and a time-guided periodic decoder.
    \item \textbf{CognTKE} \cite{CognTKE} is a cognitive-inspired framework that performs interpretable TKG extrapolation by sequentially applying a global shallow reasoner and a local deep reasoner over a TCR-Digraph.
    \item \textbf{LogCL} \cite{LogCL} is a contrastive learning framework that learns robust TKG representations. Its core mechanism is to effectively fuse local and global historical evolution information.
    \item \textbf{GPT-NeoX} \cite{GPT-NeoX} serves as a pure LLM baseline by linearizing the TKG query and its historical context into a text prompt, relying entirely on the model's in-context learning to generate the answer.
    \item \textbf{GenTKG} \cite{GenTKG} employs temporal logic rules to retrieve temporally and logically related historical facts, and then adopts a small number of instructions to fine-tune the LLM for TKGR.
    \item \textbf{CoH} \cite{lama-2-7b-CoH} is a fine-tuning framework that enhances LLMs for TKG completion by using a structure-augmented history of facts as input.
    \item \textbf{MESH} \cite{MESH} integrates structural (GCN) and semantic (LLM) information using specialized experts for historical and non-historical events.
    \item \textbf{LLM-DA} \cite{LLM-DA} allows an LLM to adapt to the evolving TKG by dynamically updating a set of logical rules that guide the model's reasoning process.
    
\end{itemize}

\section{Parameter Details}
\label{apd:parameter}
\subsection{Hyperparameter Settings}
All experiments are implemented in the PyTorch framework and conducted on a server cluster equipped with eight NVIDIA A40 GPUs (48GB VRAM each). 
To ensure the reliability and stability of the findings, all experimental results are reported as the average of multiple independent  5 runs.

For the evolving graph encoder, we follow the optimal hyperparameter configurations reported in its original implementation, setting the output embedding dimension to 200. 
The encoder is pre-trained for 500 epochs on each dataset, with its parameters subsequently frozen to support the instruction-tuning phase.

We adopt Llama3-8B\cite{llama3}, Qwen2.5-7B \cite{qwen2.5} and Mistral-7B \cite{Mistral} as the backbone LLM. Our proposed STK-Adapter incorporates three MoE modules, each comprising 4 experts with an internal hidden dimension of 8. 
To ensure computational efficiency during inference, we employ a Top-$1$ routing strategy. 
The auxiliary load-balancing loss weight is set to 0.01 to ensure expert utilization stability and $\alpha$ is set to 0.1.

The STK-Adapter is fine-tuned for 2 epochs across all datasets. We utilize the AdamW optimizer with a learning rate of $2 \times 10^{-5}$ and a weight decay of 0.01.
A linear warmup ratio of 0.01 is applied. 
To prevent gradient explosion, we apply gradient clipping with a threshold of 1.0.

To optimize memory footprint, training is performed with BF16 precision, supporting a maximum sequence length of 2048 tokens.
During inference, the beam width for beam search is set to 20. 
For the hybrid scoring strategy, the weighting coefficient $\lambda$ is empirically set to 0.1.

\begin{figure}[ht]
\centering
    \begin{minipage}[t]{1\linewidth}
    \resizebox{1\linewidth}{!}{
        \centering
        \includegraphics[width=1\linewidth]{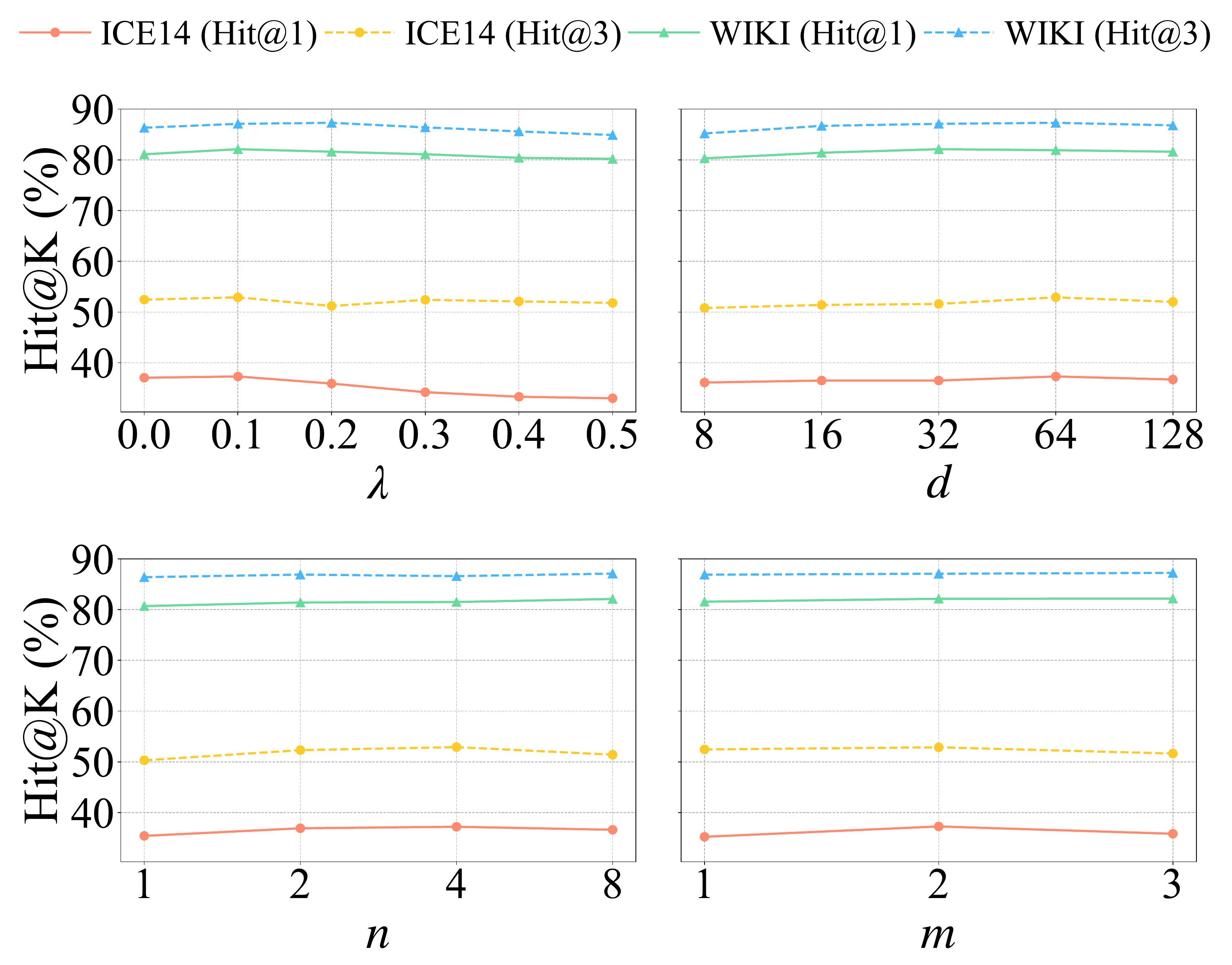}
        }
    \end{minipage}
\caption{Study on the proportion $\lambda$ of the hybrid score, the dimension $d$ of each expert, the number of experts $n$, and the depth $m$ of neighborhood sampling.}
\label{fig:parameter}
\end{figure}

\begin{figure*}[ht]
\centering
    \begin{minipage}[t]{1\linewidth}
    \resizebox{1\linewidth}{!}{
        \centering
        \includegraphics[width=1\linewidth]{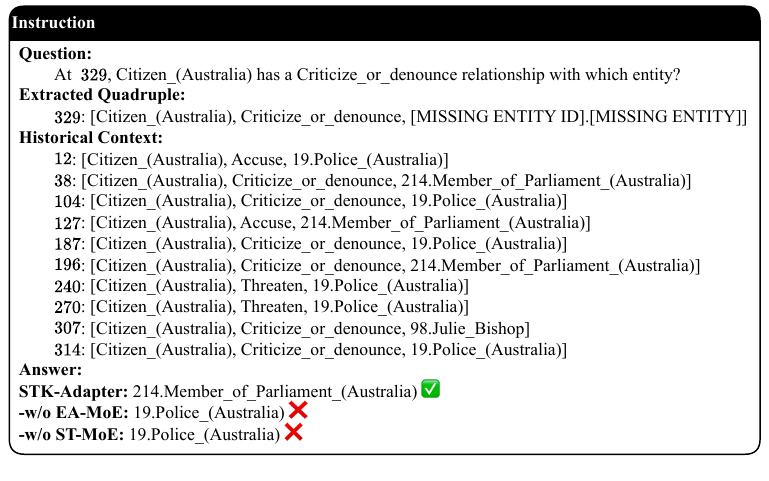}
        }
    \end{minipage}
\caption{Visualization of case study.}
\label{fig:case1}
\end{figure*}

\subsection{Parameter Sensitivity Study}
We conduct experiments on the ICE14 and WIKI datasets to explore the effects of four hyperparameters: the hybrid score proportion $\lambda$, the number of experts $n$, the dimension $d$ of each expert and the depth $m$ of neighborhood sampling. 
Figure \ref{fig:parameter} shows the results under different hyperparameter settings.
For the hybrid score proportion $\lambda$, we observe that as the weight of the graph model score increases, overall performance is relatively low when $\lambda=0.0$, peaks at $\lambda=0.1$, and then gradually declines.
This indicates that while the graph model's global ranking provides useful information, the STK-Adapter enhanced LLM is the more reliable reasoning model. 
For the dimension $d$ of each expert, changes in its value result in only minimal performance fluctuations, indicating that STK-Adapter is not highly sensitive to this parameter.
Given that a smaller $d$ reduces computational overhead while incurring only a minor performance degradation, we selected $d$=8 to balance performance and efficiency.
For the number of experts $n$, the model's performance initially improves as $n$ increases before declining.
This trend reflects that a moderate number of experts helps capture diverse patterns, while too many cause insufficient training per expert.
A similar trend is observed for the neighborhood sampling depth $m$, where appropriate sampling depth provides essential contextual information, while excessive depth may introduce irrelevant information.

\section{Analysis of Layer-wise Injection}
To evaluate the impact of layer-wise STK-Adapter injection, we conduct controlled experiments on ICE14, where STK-Adapter is randomly injected into a single layer, 20\% of layers, and 50\% of layers.
As shown in Table \ref{tab:inject}, performance consistently improves as the number of injected layers increases. 
This confirms that maximizing the number of injected layers is beneficial.

\begin{table}[ht]
\centering
\caption{Impact of Injection Coverage on Performance.}
\label{tab:inject}
\setlength\tabcolsep{8pt} 
\small
\begin{tabular}{lccc}
\toprule
& \multicolumn{3}{c}{\textbf{ICE14}} \\
\cmidrule(lr){2-4}
\textbf{Coverage} & \textbf{Hit@1} & \textbf{Hit@3} & \textbf{Hit@10} \\
\midrule
100\% & 37.26 & 52.86 & 67.57  \\
50\% & 33.93 & 48.18 & 63.42  \\
20\% & 30.37 & 44.21 & 62.31 \\
Single Layer & 26.26 & 41.78 & 58.66 \\
\bottomrule
\end{tabular}
\end{table}

\section{Case Study}
\label{apd:case}

We analyze a representative case on the ICE14 dataset. 
As illustrated in Figure \ref{fig:case1} at timestamp $t=329$, an entity Citizen\_Australia performs a Criticize\_or\_denounce relation toward a missing entity.
Despite the prevalence of high-frequency entities (e.g., Police) in the historical event chain, which may introduce frequency bias,
STK-Adapter is still able to accurately predict the missing entity as Member\_of\_Parliament\_Australia, demonstrating its effectiveness in integrating evolving structural and temporal semantic information.

The variant $\textit{-w/o EA-MoE}$ is highly susceptible to frequency bias. 
In the absence of temporal event semantic dependencies, LLMs primarily rely on text co-occurrence frequencies for prediction.
Consequently, this leads to a propensity to erroneously predict Police\_(Australia) as the target entity, because it appears most frequently in the historical event chains.

The variant $\textit{-w/o ST-MoE}$ suffers from instability in semantic decision-making.
Although it can perceive latent correlations between events, the absence of explicit grounding in temporal graph structures makes it difficult to distinguish the object-specific nuances between Criticize and Accuse within a temporal context.
Consequently, the model vacillates among various political or social entities, failing to yield deterministic predictions.

\section{LLM Usage}
LLMs were utilized solely to assist with the writing and linguistic refinement of this manuscript. 
In particular, the LLM was employed to improve clarity, readability, and overall textual coherence, including tasks such as grammar checking, sentence rephrasing, and stylistic polishing.

The LLM did not participate in the formulation of research ideas, methodological design, experimental setup, data analysis, or interpretation of results. 
All scientific contributions, including the research motivation, model design, experimental evaluation, and conclusions, were entirely developed by the authors.

The authors take full responsibility for the content of the manuscript, including any text that was refined with the assistance of the LLM.
All uses of the LLM were conducted in accordance with ethical guidelines, and the authors ensured that its assistance did not introduce plagiarism, fabricated content, or any form of scientific misconduct.

\end{document}